\documentclass[fleqn,10pt]{wlscirep}
\usepackage[utf8]{inputenc}
\usepackage[T1]{fontenc}
\usepackage{bm}
\usepackage{amsmath}
\usepackage{amssymb}
\usepackage{ragged2e}
\usepackage[user,titleref]{zref}

\graphicspath{ {./figures/} }

\title{Nanoscale spin ordering and spin screening effects in tunnel ferromagnetic Josephson junctions}

\author[1,2,*]{Roberta Satariano}
\author[3]{Anatoly Fjodorovich Volkov}
\author[1,2]{Halima Giovanna Ahmad}
\author[1]{ Luigi Di Palma}
\author[1,2]{Raffaella Ferraiuolo}
\author[4]{Antonio Vettoliere}
\author[4]{Carmine Granata}
\author[1,2]{Domenico Montemurro}
\author[1,2]{Loredana Parlato}
\author[1,2]{Giovanni Piero Pepe}
\author[1]{Francesco Tafuri}
\author[1,2]{Giovanni Ausanio}
\author[5,2,**]{Davide Massarotti}

\affil[1] {Dipartimento di Fisica Ettore Pancini, Università degli Studi di Napoli Federico II, c/o Complesso Monte Sant’Angelo, via Cinthia 26, I-80126 Napoli, Italy}
\affil[2]{Consiglio Nazionale delle Ricerche - SPIN, c/o Complesso Monte Sant’Angelo, via Cinthia 26, I-80126 Napoli, Italy}
\affil[3]{Institut f\"ur Theoretische Physik III, Ruhr-Universit\"at Bochum, Universit\"atsstraße 150, D-44780 Bochum, Germany}
\affil[4]{Consiglio Nazionale delle Ricerche—ISASI, Via Campi Flegrei 34, I-80078 Pozzuoli, Italy}
\affil[5]{Dipartimento di Ingegneria Elettrica e delle Tecnologie dell’Informazione, Università degli Studi di Napoli Federico II, Via Claudio 21, 80125 Napoli, Italy}
\affil[*]{e-mail: roberta.satariano@unina.it}
\affil[**]{e-mail: davide.massarotti@unina.it}
\begin{abstract} 
Magnetic Josephson junctions (MJJs) have emerged as a prominent playground to explore the interplay between superconductivity and ferromagnetism. A series of fascinating experiments have revealed striking phenomena at the Superconductor/Ferromagnet (S/F) interface, pointing to tunable phase transitions and to the generation of unconventional spin-triplet correlations. 
Here, we show that the Josephson effect, being sensitive to phase space variation on the nanoscale, allows a direct observation of the spin polarization of the S/F interface. By measuring the temperature dependence of the Josephson magnetic field patterns of tunnel MJJs with strong and thin F-layer, we demonstrate an induced nanoscale spin order in S along the superconducting coherence length at S/F interface, i.e., the inverse proximity effect, with the first evidence of full spin screening at very low temperatures, as expected by the theory. A comprehensive phase diagram for spin nanoscale ordering regimes at S/F interfaces in MJJs has been derived in terms of the magnetic moment induced in the S-layer. Our findings contribute to drive the design and the tailoring of S/F interfaces also in view of potential applications in quantum computing.
\end{abstract}

\begin{document}

\flushbottom
\maketitle
\thispagestyle{empty}

\section*{Introduction}

Superconductors (S) and Ferromagnets (F) heterostructures are a unique platform where antagonistic correlations, namely the exchange interaction and the superconducting phase coherence, combine \cite{ BUDZIN2005,BERGERET2005_REVIEW,Volkov_PRB2001,GOLUBOV2004}. 
The two competing orders not only generate unconventional types of Cooper pairs or ordered phases, but also offer novel paradigms to realize tunable Josephson junctions (JJs). As matter of fact, hybrid JJs  integrating superconductors and exotic barriers go beyond combining the physics of their components \cite{Supercurrent_in_van_der_Waals_Josephson_junction,Van_der_Waals_ferromagnetic_Josephson_junctions}, but their capabilities of transferring and merging different orders led to novel physics and functionalities in JJs, including their more recent and advanced development, the superconducting qubit \cite{ Kawabata2006,FEOFANOV,Massarotti2015,Ahmad_22PRB,Graphene2018,Graphene_QED,van_der_Waals_technology2019,Semiconductor-Nanowire-Based1-qubit,MICROWAWE_QUANTUM_CIRCUIT_SEMICONDUCTING_NANOWIRE,Ballistic_Majorana_nanowire_devices}.  The ferro-transmon \cite{Ahmad_22PRB,ferro-trasmon_path} and ferro-gatemon \cite{Semiconductor-Nanowire-Based1-qubit,ferro-gatemon_supercurrent_reversal} are in this respect two emerging promising examples, which exploit the unconventional phenomena occurring at S/F interface in magnetic JJs (MJJs) in view of innovative quantum bits.

\begin{figure} [h]
   
    \includegraphics[width = 17.5 cm]{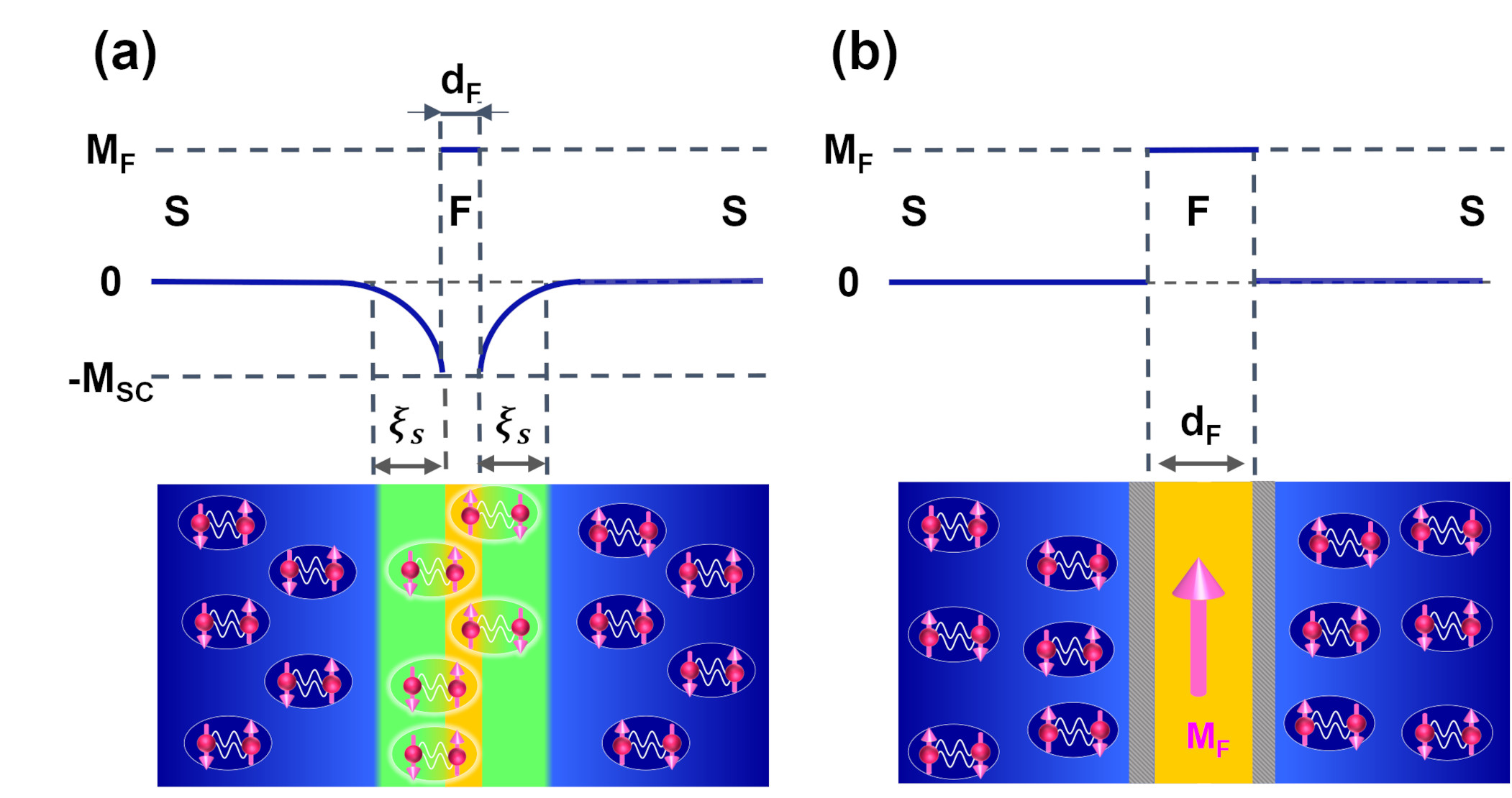}   
    \caption{ \textbf{Inverse proximity effect (IPE) in magnetic Josephson junctions (MJJs).} Sketch of the spin polarization at the superconductor (S)/ ferromagnet (F) interface in MJJs in the diffusive limit with a homogeneous ferromagnet. \textbf{a)} The electrons of the Cooper pairs at the S/F interface with the spin aligned along the exchange field penetrate into the F-layer, while the electrons with the opposite spin tend to stay in S. As a result, the surface of the S layer down a depth of the superconducting coherence length $\xi_{\mathrm{s}}$ acquires a net  magnetization \textbf{M}\(_{\mathrm{SC}}\) with opposite direction to the F-magnetization \textbf{M}\(_{\mathrm{F}}\). \textbf{b)} In MJJs with large thickness of a weak ferromagnet and low-transparency of the S/F interface, the leakage of the ferromagnetic order into the superconductor is prevented or weakly induced. The profile of the magnetization is depicted as a function of the distance from the S/F interface (blue line).}
    \label{fig:sketch_proximity_effect}
\end{figure}

A striking example of a more extended order is the inverse proximity effect (IPE), i.e., the transfer of a ferromagnetic order into a superconductor from the S/F interface \cite{BERGERET2004,Volkov_EPL_2004,BERGERET2005_REVIEW,IPE_BERGERET_PRB72,Volkov_PRB_2006, Volkov2019,Yagovtsev_2021}, which adds to the standard proximity effect, i.e., the influence of S in F \cite{BUDZIN2005,BERGERET2005_REVIEW,Volkov_PRB2001,GOLUBOV2004}. More precisely, the electrons of the Cooper pairs with the spin aligned along the exchange field can easily penetrate the F-layer, while the electrons with the opposite spin tend to stay in S \cite{BERGERET2004,Volkov_EPL_2004,BERGERET2005_REVIEW,IPE_BERGERET_PRB72,Volkov_PRB_2006, Volkov2019,Yagovtsev_2021}. As a result, the surface of the S-layer down to a depth of the order of the Cooper pair size, i.e., the superconducting coherence length \(\xi_{\mathrm{s}}\), acquires a net magnetization \(\mathbf{{M_{\mathrm{SC}}}}\) with opposite direction to the F-magnetization \(\mathbf{{M_{\mathrm{F}}}}\), which can even compensate its magnetic moment \cite{Volkov_EPL_2004,BERGERET2005_REVIEW,Volkov_PRB_2006, Volkov2019} (Fig. \ref{fig:sketch_proximity_effect}a). This effect is not universal\cite{RefereeA_PhysRevB.66.224516}: inhomogeneous ferromagnetic textures can invert the sign of the proximity-induced magnetization in the superconductor \cite{RefereeA_PhysRevB.79.054523}, while in the ballistic limit, the induced magnetization changes sign in space so that the anti-screening effect may take place \cite{RefereeA_PhysRevB.66.224516, IPE_BERGERET_PRB72, Volkov_PRB_2006}. The system studied in our work corresponds to a diffusive case with a homogeneous ferromagnet and should fully fall in the regime where the screening effect is expected. 
So far, several attempts to observe the IPE have been sought in many different S/F proximity-coupled systems, but its observation has been quite elusive since there are very few techniques able to probe the magnetic fields at nanoscale \cite{Flokstra2021, Khaydukov2002}. Indirect evidence of the induced magnetization \textbf{M}\(_{\mathrm{SC}}\) has emerged from the measurements of S/F thin films across the superconducting critical temperature \(T_{\mathrm{c}}\) \cite{XIA2009,Salikhov2009}.  Nevertheless, discriminating the spin polarization phenomena from the Meissner expulsion in response to the vector potential at S/F interface has been proved controversial \cite{Flokstra2016}, while the saturation of the induced magnetization at low temperatures, as expected for high-transparent S/F interfaces, has never been reported\cite{XIA2009}, thus the comparison with theoretical models is still under debate.

Over the last decades, advances in the fabrication and design of MJJs with a rich variety of materials, geometries and  layouts have established a powerful platform to reveal new physical phenomena at the S/F interface. For instance, direct evidence of  0-\(\mathrm{\pi}\) phase transition has been provided in SFS JJs with large thickness of weak ferromagnets \cite{Ryazanov2001,sellier2003,Frolov2004,Sellier2004,kontos2002,Blum2002,Oboznov2006,WEIDES2006,Bannykh_0_pi_transition_2009,Khaire2009,GLICK2017,o-pi_yamashita_2017},  while spin-triplet pairing has been generated by introducing some magnetic non-collinearity, resulting in anomalously long scaling lengths \cite{ECHRING2015,LINDER2015,Banerjee2014_bis,Banerjee2014,spin_triplet_new_robinson_science,spin_triplet_new_ref_robinson2011,spin_triplet_new_ref_khaire2010,spin_triplet_new_leksin2012,spin_triplet_new_anwar2012,spin_triplet_new_klose2012}. 
However, such MJJs are not suitable for detecting the spin polarization of the S/F interface: in SFS JJs with large thickness of the F-layer, usually characterized by relatively low exchange field, or in MJJs containing complex ferromagnetic/normal metal multilayer the nanoscale spin ordering is only weakly induced (Fig. \ref{fig:sketch_proximity_effect}b). 

In this work, by exploiting SIsFS tunnel junctions with strong and thin F-barriers, we define the conditions to unambiguously distinguish different spin screening regimes and tune the alignment of spins at the S/F interface as a function of the temperature. The nanoscale spin arrangement manifests itself directly in the magnetic dependence of the Josephson critical current. The hallmarks of the strong polarization limit are: i) the lack of hysteresis of the magnetic field patterns and ii) the broadening of their central peak \cite{DAHIR2019}. By measuring the \(I_{\mathrm{c}}\) curves as a function of the temperature \textit{T} down to \textit{T} = 10 mK, we have used the temperature as an external knob to control this effect and to clearly identify the spin polarization of the Cooper pairs \cite{DAHIR2019}. The temperature behaviour of the magnetic field patterns is consistent with the theoretical predictions, thus confirming the crucial role of the superconducting gap \(\Delta\), the magnetic exchange energy \textit{J} and the transparency of the S/F interface\cite{BERGERET2004,Volkov2019}.

\section*{Results} \label{RE}
\subsubsection*{Transport properties of SIsFS JJs \label{section:transport_properties}}
Our SIsFS JJs are based on a standard Nb trilayer technology and exploit a Ni$_{80}$Fe$_{20}$ alloy (Permalloy: Py) as F-barrier \cite{JAP}. The sketch of the SIsFS JJs [Nb (200 nm)/ Al-AlO$_{\mathrm{x}}$ (7 nm)/ Nb (30 nm)/ Py (3 nm)/ Nb (400 nm)] and the reference system axis are reported in Fig. \ref{fig:sketch_and_IV}a. Details on the fabrication process can be found in the Section \ztitleref{methods}. Fig. \ref{fig:sketch_and_IV}b shows the current-voltage (I-V) characteristics measured at \textit{T} = 10 mK for circular SIsFS JJs with radius \( R = 2 \ \mu \mathrm{m} \) (black curve) and with \( R = 1.5 \  \mu\mathrm{m} \) (red curve), respectively. The I-V characteristic of non-magnetic SIsS JJ with \(R = 2.5 \ \mu\mathrm{m} \) is reported as a term of comparison (blue curve).  Properties of SIsFS JJs can be discussed in the framework of the theoretical model proposed in Refs. [\citenum{BAKURSKIY2013,BAKURSKIY2013_PRB}].  Basically, different transport regimes can be distinguished by comparing the thickness of the intermediate superconducting layer \(d_{\mathrm{s}}\) with the critical thickness \textit{d}\(_{\mathrm{sc}}\), i.e., the minimal thickness of the s layer in a sF bilayer above which superconductivity still exists at a certain temperature. If \(d_{\mathrm{s}}\) is sufficiently larger than \textit{d}\(_{\mathrm{sc}}\), the pair potential \(\Delta\)\ in the s layer is close to that of the bulk material and the SIsFS structure can be considered as a series of a tunnel SIs JJ and a ferromagnetic sFS JJ\cite{BAKURSKIY2013,BAKURSKIY2013_PRB}. For small F-thickness \textit{d}\(_{\mathrm{F}}\), because of the metallic nature of standard F barrier and resulting higher barrier transparency, the critical current of the SIs side is expected to be much smaller than the one of sFS side. For instance, at helium-liquid temperature SIS JJs based on Nb trilayer technology have critical current density values ranging from tens A cm$^{-2}$ ( Ref. [\citenum{Castellano_2006}]) to k Acm$^{-2}$ (Ref. [\citenum{Yohannes2005}]), some orders of magnitude less than the value commonly measured for SFS \cite{Kapran2021,Robinson2006,Bell2005}.  
Therefore, since \(I_{\mathrm{1c}} \ll I_{\mathrm{2c}}\), where \(I_{\mathrm{1c}}\) and \(I_{\mathrm{2c}}\) are the critical current of the SIs and sFS side, respectively,\ the I-V curve of the overall SIsFS device is determined by its SIs part and the critical current-normal resistance product \textit{I}\(_{\mathrm{c}}\)\textit{R}\(_{\mathrm{N}}\) can reach its maximum value corresponding to a standard SIS JJ\cite{BAKURSKIY2013,BAKURSKIY2013_PRB}. This is precisely the regime in which our SIsFS JJs fall. Since the s interlayer in our JJs (\( \mathit{d_{\mathrm{s}}} = 30 \ \textrm{nm}\)) is sufficiently thicker than the superconducting coherence length \(\xi_{\mathrm{s}}\) ( $\sim$10\ nm), the SIs JJ with the smaller critical current sets the behavior of the overall structure resulting in \(j_{\mathrm{c}}\)  value of the order of 50 A cm$^{-2}$ and in  \textit{I}\(_{\mathrm{c}}\)\textit{R}\(_{\mathrm{N}}\) values of 1 mV at \textit{T} = 10 mK, as the standard tunnel SIsS junctions \cite{BAKURSKIY2013}. The latter values are reduced by only 20$\%$ with respect to the reference SIsS JJ in the whole temperature range analyzed in this work, from 10 mK up to 6 K. Details on the method by which the coherence lengths have been determined can be found in the Supplementary Note 1, while the I-V characteristics as a function of the temperature are reported in Supplementary Figure 1. 
The high-quality tunnel behaviour of the SIsFS JJs is evident also from the shape of the subgap branch, which  does not show any evident deviation from the reference SIsS JJ (Fig. \ref{fig:sketch_and_IV}b).  
We have thus fitted the I-V characteristics with the Tunnel Junction Microscopic (TJM) model \cite{BARONE1982} (see Supplementary Figure 2), which is a well-established technique to analyze the electrodynamics of SIS tunnel junctions \cite{Ahmad_PRA20}. In conclusions, the experimental evidence indicates that the superconductivity in the s-interlayer is not suppressed and thus the SIsFS JJs behave as a serial connection of a SIs and an sFS JJ with the transport properties dominated by the SIs part. For a detailed account of the electrodynamics parameters of these junctions at \textit{T}  = 10 mK, we refer to Supplementary Table 1.         

\subsubsection*{Magnetic field patterns}
\begin{figure} [h]
   
    \includegraphics[width = 17.5 cm]{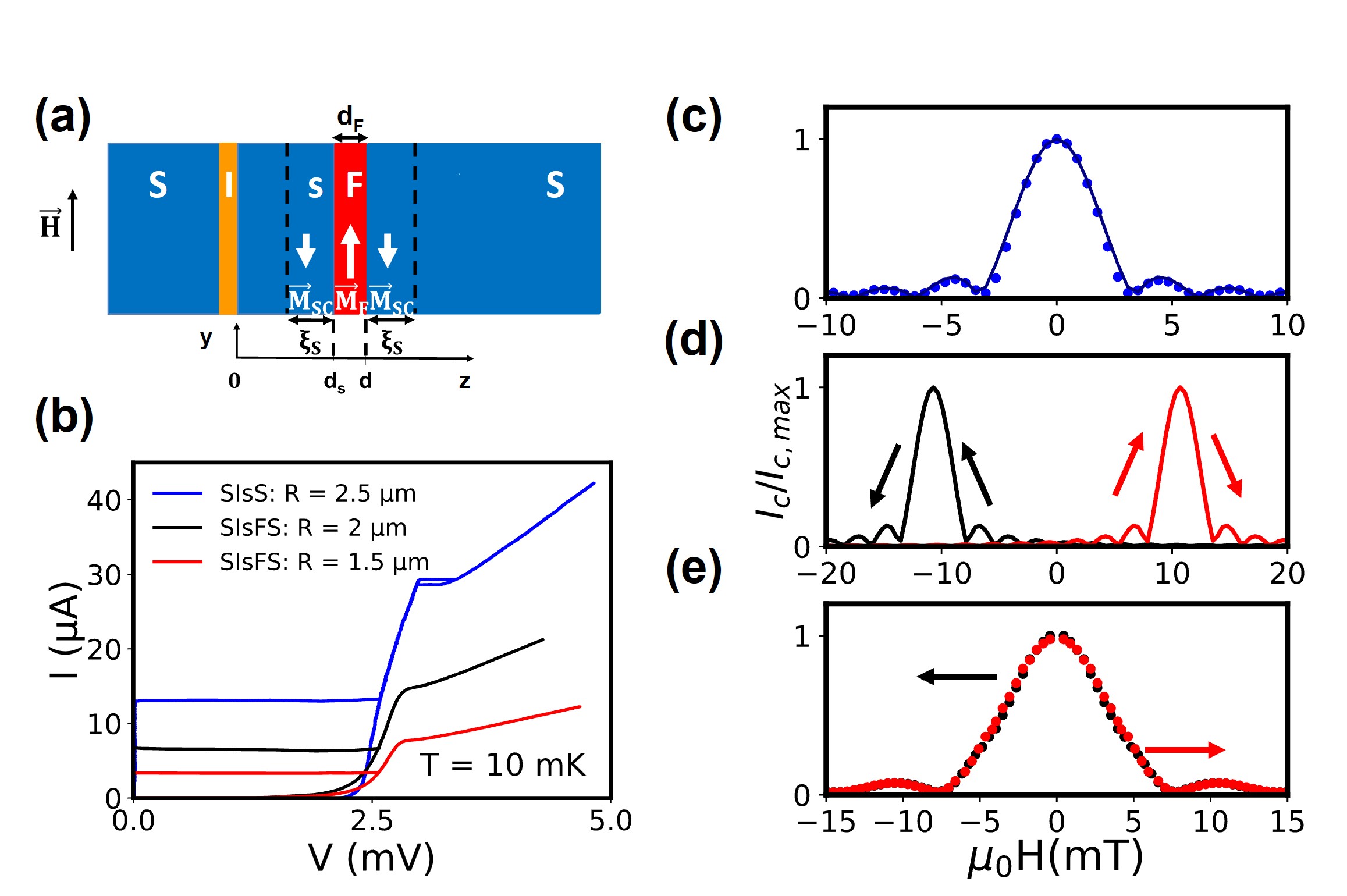}   
         \caption{ \textbf{Superconductor/ Insulator/ thin superconductor/ Ferromagnet/ Superconductor Josephson junctions (SIsFS JJs) and comparison with the non-magnetic junctions.}  \textbf{a)} Sketch of the SIsFS JJs and reference system. \textbf{b)} Current-voltage (I-V) characteristics for a circular SIsFS JJ with radius \( R = 2  \ \mu\mathrm{m} \) (black curve) and with \( R = 1.5 \  \mu\mathrm{m} \)  (red curve), and for a SIsS JJ with \( R = 2.5 \ \mu\mathrm{m} \) (blue curve). \textbf{c)} Normalized \textit{I}\(_{\mathrm{c}}\) as a function of the magnetic field \textit{H} for a circular SIsS JJ with \( R = 1.5 \  \mu\mathrm{m} \). The solid blue line indicates the Airy pattern fit. \textbf{d)} Reconstructed \(I_{\mathrm{c}}\) curves in a SIsFS JJ with \( R = 1.5 \  \mu\mathrm{m} \) in absence of  spin polarization by considering an Airy pattern, the flux expression in Eq. (\ref{eq1}) and the measured hysteresis loop for a Py micrometer dot reported in Supplementary Figure 3a. \textbf{e)} Measured \(I_{\mathrm{c}}\) curves of a circular SIsFS JJ with \( R = 1.5 \  \mu\mathrm{m} \). All experimental data are collected at \textit{T }= 10 mK. In our experimental setup, the current and voltage are affected by errors of 1\% and 2\%, respectively\cite{Ahmad_PRA20}. In both panels d and e, the black and red curves are the magnetic patterns in the downward and upward direction of the magnetic field, respectively. The arrows indicate the sweeping field directions.} \label{fig:sketch_and_IV}
\end{figure}

\begin{figure} [h]
   
    \includegraphics[width = 17.5 cm]{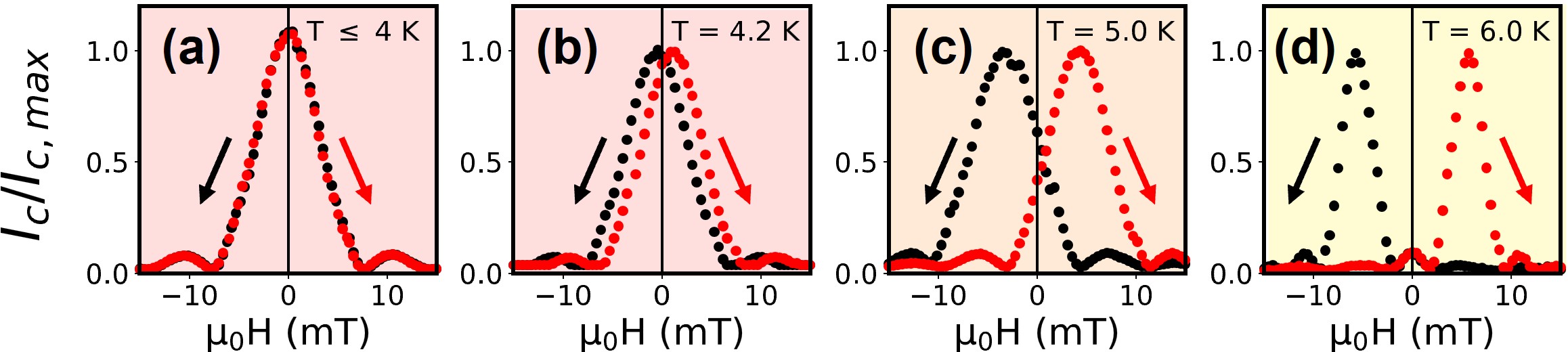}   
\caption{ \textbf{Temperature dependence of the magnetic field pattern for the SIsFS Josephson junction with \(\boldsymbol{ R = 1.5 \  \mu}\)m.} We measured the \(I_{\mathrm{c}}\) curves by sweeping the field in the range (-15, 15) mT at different temperatures \textit{T}: \textbf{a)} \textit{T} = 0.01 K, 1 K, 2 K, 3 K and 4 K, \textbf{b)} \textit{T} = 4.2 K, \textbf{c)} \textit{T} = 5 K, and \textbf{d)} \textit{T} = 6 K. In the panel a, for \textit{T} \(\leq\) 4 K, the measurements do not show any deviation within this temperature range. The error bar on each measured \textit{I}\(_{\mathrm{c}}\) point is of the order of  1$\%$ (Ref. [\citenum{Ahmad_22}]). The black and red curves are the magnetic pattern in the downward and upward direction of the magnetic field, respectively. The arrows indicate the sweeping field direction.}
    \label{Temperature_dependence_no_shift}
\end{figure}

In Fig. \ref{fig:sketch_and_IV}c,  the magnetic field pattern at \textit{T }= 10 mK of non-magnetic SIsS JJ with \( R = 1.5 \  \mu\mathrm{m} \) is reported. 
The magnetic pattern is consistent with the expected Airy pattern \cite{BARONE1982}: \(
I_{\mathrm{c}}/I_{\mathrm{c,max}} = 
\left |\frac {2j_{1} \left (\frac{ {\pi\Phi}} {\Phi_{0} } \right) } {\frac{\pi\Phi } {\Phi_{0} } }\right |,\)
where  \textit{j$_{1}$} is a Bessel function of the first kind and \begin{math}  \Phi_{0} = \frac{h}{2e}\end{math} is the magnetic quantum flux. The magnetic flux through the junction is \begin{math} \Phi \  = \ \mu_{\mathrm{0}} H 2 R   \ (d_{\mathrm{I}} +d_{\mathrm{s}} + 2\lambda_{\mathrm{L}})  \end{math}, where \(d_{\mathrm{I}}\) and \(d_{\mathrm{s}}\) are the thicknesses of the I and s layers, respectively, and  \(\lambda_{\mathrm{L}}\) is the London penetration depth. The fitted values \(R = 1.52 \pm 0.02  \ \mu\mathrm{m}\) and \(\lambda_{\mathrm{L}}\) = 120 $\pm$ 20 nm are in agreement with nominal junction dimensions and expected \(\lambda_{\mathrm{L}}\) for Nb \cite{Kapran2021,Khaire2009}.  
If the s-interlayer is too thin to screen the magnetic fields by Meissner effect \begin{math} (d_{\mathrm{s}} < \lambda_{\mathrm{L}})\end{math}, the in-plane magnetization magnetization of the F-layer \(\mathbf{{M_{\mathrm{F}}}}\) contributes to the total magnetic flux through the junction \cite{BAKURSKIY2013,Golovchanskiy2016}:
\begin{eqnarray}\label{eq1}
\Phi = \mu_{\mathrm{0}}H2Rd_{\mathrm{m}}  + \mu_{\mathrm{0}}M_{\mathrm{F}}2Rd_{\mathrm{F}}, 
\end{eqnarray}where the thickness of the material penetrated by the applied field is \(d_{\mathrm{m}} = 2 \lambda_{\mathrm{L}} + d_{\mathrm{s}} + d_{\mathrm{F}} + d_{\mathrm{I}} \), with \(d_{\mathrm{F}}\) the thickness of the intermediate F-layer \cite{LARKIN2012}.  As a result, even if the SIsFS JJs act as a serial connection between a SIs and a sFS JJ at very low temperatures,  circular SIsFS JJs with \begin{math} d_{\mathrm{s}} < \lambda_{\mathrm{L}}\end{math} behave as a single junction with respect to an external field \textbf{H} and present an Airy-like pattern shifted from
zero field in agreement with the magnetic hysteresis of the F-layer \cite{Vettoliere22}.  In particular, the maximum of the \(I_{\mathrm{c}}\) curves corresponds to a zero total magnetic flux across the MJJ. If the coercive field is large enough and the F-magnetization \(\mathbf{{M_{\mathrm{F}}}}\) follows a Stoner–Wohlfarth single-domain behavior \cite{Cowburn_2000}, we expect a standard Airy pattern with a shift in field by: \(\pm \mu_{\mathrm{0}}H\mathrm{_{shift}} = \mp   \mu_{\mathrm{0}}M_{\mathrm{F}}d_{\mathrm{F}} / d_{\mathrm{m}}\), where  \(\mu_{\mathrm{0}}M_{\mathrm{F}}\) corresponds to the saturation magnetization. If we consider \(\mu_{\mathrm{0}}M_{\mathrm{F}} = 1\) T for the Py layer  \cite{JAP,Satariano} and junction dimensions of the JJ in Fig. \ref{fig:sketch_and_IV}c, we should expect a shift in field in the upward direction at  \(\mu_{\mathrm{0}}H\mathrm{_{shift}} \sim \ 11 \) mT. However, this situation represents an upper limit for the shift of a magnetic field pattern in a MJJ: the rotation of the magnetization in the domains or the domain wall motions can result into narrow central peaks and displacements of the magnetic field patterns \cite{BOL'GINOV2012,Khaire2009}. If we reconstruct the Airy pattern by taking into account the flux expression in Eq. (\ref{eq1}) and the hysteresis loop for a micrometer Py dot in Supplementary Figure 3a, we expect a shift at \( \mu_{\mathrm{0}}H\mathrm{_{shift}} \sim \ 5 \) mT and an almost unchanged width (Supplementary Figure 3b), as confirmed by the transport measurements reported in Fig. 3d.
\newline
 In contrast, in the experimental patterns of the SIsFS JJs at \textit{T }= 10 mK  we observe two main anomalies: the widening of the central peak of about a factor 2.5 and the lack of hysteresis (Fig. \ref{fig:sketch_and_IV}e).  In Fig. \ref{Temperature_dependence_no_shift}, we show the magnetic field patterns for a SIsFS JJ with \( R = 1.5 \  \mu\mathrm{m} \) measured as a function of the temperature. Zero-shifted curves are observed below \textit{T}= 4 K, while above \textit{T} = 4 K the ordinary hysteresis is recovered, as in Fig. \ref{fig:sketch_and_IV}d. As the temperature increases, the shift in field increases. 
In this narrow range of temperature so far below the Curie temperature of Py and for the size of our nanomagnet, changes of the magnetization curve of the F-layer are negligible (see Supplementary Figure 4 and the Supplementary Note 2). Moreover, the absence of hysteresis of the \(I_{\mathrm{c}} (H)\) curves in Fig. \ref{Temperature_dependence_no_shift}a cannot be related to a vortex state of the F layer at zero field since we observe a net remanence in our Py micrometer dots (compare Supplementary Figure 4b with Fig. 13a in Ref. [\citenum{Cowburn_2000}]).
In addition, at \textit{T} = 6 K the width of the central peak is halved, while a reduction of only 20\% is expected if we consider the temperature dependence of \(\lambda_{\mathrm{L}}(T)\). These experimental observations have been consistently measured in different junctions on different samples with the same geometry reported in Fig. \ref{fig:sketch_and_IV}e. This unconventional phenomenology of the magnetic field patterns can be discussed in the frame of the IPE.
\section*{Discussion} \label{DIS}
\begin{figure} [h]
   \centering
    \includegraphics[width = 9 cm]{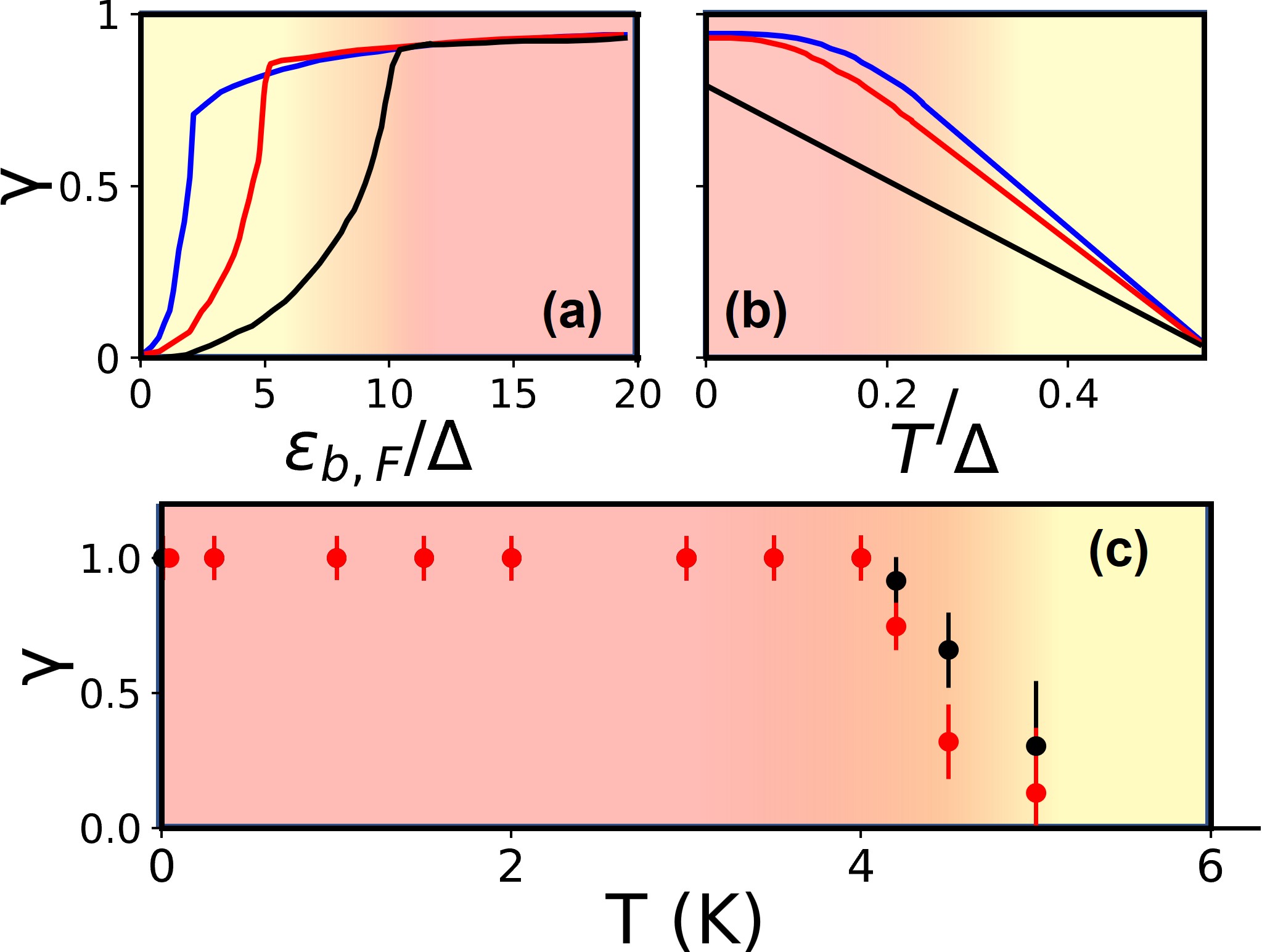}   
 \caption{ \textbf{Theoretical dependence of the induced magnetic moment in the superconducting layer and comparison with the experimental data.} \textbf{a) }Theoretical dependence of  $\gamma$, i.e., the magnetic moment of the S-layers normalized to the F-layer in absolute value, on the characteristic energy \(\varepsilon_{\mathrm{b,F}} = \hslash D_{\mathrm{F}}/({R_{\mathrm{b}}\sigma_{\mathrm{F}}d_{\mathrm{F}}})\), where \textit{D}\(_{\mathrm{F}}\) is the diffusion coefficient of the F-layer, \(R_{\mathrm{b}}\)  is the S/F interface resistance per unit area, and \textit{$\sigma_{\mathrm{F}}$} is the F-conductivity. The calculations have been obtained at the normalized temperature \(T/ \Delta = 0.01\) for  \(J / \Delta  = 10,  \ 5,  \ 2\) (black, red, blue curve, respectively), where \textit{J} is the exchange energy of the F-layer and \textit{$\Delta$} is the superconducting gap.  \textbf{b)} Theoretical dependence of $\gamma$ on the reduced temperature \begin{math} T/ \Delta \end{math}  for  \(J / \Delta  = 10\), and for different value of \( \varepsilon_{\mathrm{b,F}} / \Delta(0) = 10, \ 15, \ 20\) (black, red, blue curve, respectively). \textbf{c)} Experimental temperature dependence of \(\gamma\) for the set of measurements in Fig. \ref{Temperature_dependence_no_shift}: the black and red points have been derived for the downward and upward magnetic field curves, respectively. The error bars on $\gamma$ are calculated by propagating the errors on the magnetic thickness \(d_{\mathrm{m}}\) and on the magnetic field corresponding to the maximum of \textit{I}\(_{\mathrm{c}}\). In all the panels, the pink regions indicate the regime of a strong polarization, while the light yellow ones indicate that the IPE is weakly induced. The background colours in panel a and b have been chosen with respect to the black and blue curve, respectively.}
    \label{temperature_dependence_induced_magnetization}
\end{figure}

The effect of the spin polarization of the Cooper pairs on the Fraunhofer pattern in SFS JJs has been discussed in Ref.  [\citenum{DAHIR2019}] by using a microscopic model in the dirty limit. The dimensionless magnetic moment \begin{math}  \gamma \ = \ \left | \frac{\mathcal{M}_{\mathrm{SC}} }{\mathcal{M}_{\mathrm{F}}} \right | \end{math}, i.e., the ratio between the overall magnetic moment induced in the adjacent superconductors  \(\mathcal{M}_{\mathrm{SC}} = {M_{\mathrm{SC}}}2\xi_{\mathrm{s}}\) and the magnetic moment of the F-layer \begin{math}\mathcal{M}_{\mathrm{F}} = M_{\mathrm{F}}d_{\mathrm{F}}\end{math}, is the key parameter to quantify the spin screening at the S/F interface. \(\gamma\) can be expressed in terms of the exchange energy \textit{J}, the superconducting gap \textit{$\Delta$} and the transparency of the S/F interface through the parameter:  
\(\varepsilon_{\mathrm{b,F}} = \hslash D_{\mathrm{F}}/({R_{\mathrm{b}}\sigma_{\mathrm{F}}d_{\mathrm{F}}})\), where \textit{D}\(_{\mathrm{F}}\) is the diffusion coefficient of the F-layer, \(R_{\mathrm{b}}\)  is the S/F interface resistance per unit area, and \textit{$\sigma_{\mathrm{F}}$} is the F-conductivity.  As addressed in the Supplementary Note 3, we have evaluated the dependence of $\gamma$ on  \( \varepsilon_{\mathrm{b,F}} / \Delta\) at \(T/ \Delta = 0.01\) for  \(J / \Delta  = 10,  \ 5,  \ 2\) (black, red, blue curve, respectively, in Fig. \ref{temperature_dependence_induced_magnetization}a ).  The kink at \(J \cong \varepsilon_{\mathrm{b,F}} \) marks the crossover between a weak magnetic order regime, occurring at very low temperatures  at \(\varepsilon_{\mathrm{b,F}} \leq J\), i.e., for large thickness of the F layer or for poor S/F interface, and an almost full spin screening  regime in the limit \(\varepsilon_{\mathrm{b,F}} \geq J\). As \(\varepsilon_{\mathrm{b,F}} \) increases, the latter limit is observed for a larger temperature range (Fig. \ref{temperature_dependence_induced_magnetization}b). 
In Fig. \ref{temperature_dependence_induced_magnetization}b, we report the temperature dependence of \(\gamma \) for  \(J / \Delta  = 10\), which is suggested by the parameters of the Nb/Py system under investigation  \cite{Robinson2006}, and for \( \varepsilon_{\mathrm{b,F}} / \Delta = 10, \ 15, \ 20\) (black, red, blue curve, respectively), while further detail on the dependence of $\gamma$ on  \( J / \Delta\) can be found in the Supplementary Figure  5.  

Under the conditions mentioned above, the \(I_{\mathrm{c}}(H)\) curves show clear signatures of spin screening effects. Indeed, in a standard SFS JJ, Eq. (\ref{eq1})  turns into \cite{DAHIR2019}:
\begin{equation}\label{flux_dahir}
\Phi = \mu_{\mathrm{0}}H2Rd_{\mathrm{m}}  + \mu_{\mathrm{0}}M_{\mathrm{F}}2Rd_{\mathrm{F}}  ( 1 - \gamma),
\end{equation}
where \( d_{\mathrm{m}} = 2 \lambda_{\mathrm{L}} + d_{\mathrm{F}}, \) in this case. If \(d_{\mathrm{F}}\) is much less than \(\xi_{\mathrm{s}}\), the flux due to the S-magnetization can be comparable to the one due to the F-magnetization \((\gamma \simeq  1)\), resulting in a zero-net shift of the magnetic field pattern\cite{DAHIR2019}. Eq. (\ref{flux_dahir}) can be extended to a stacked multilayer structure. The relation between an applied magnetic field \textbf{H} and the in-plane gradient of the phase difference across the SIs junction  $\nabla \varphi $ and across the sFS JJ  $\nabla \psi $ allows to evaluate the actual magnetic $\Phi$ through the junction layout.  The problem is simplified if we consider  that our SIsFS JJs behave as a serial connection of a SIs and a sFS JJ. The phase difference $\varphi$ is thus coupled to $\psi$ via the relation: \( I_{\mathrm{1c}}\sin \varphi = I_{\mathrm{2c}}\sin \psi,\) where again \(I_{\mathrm{1c}}\) and \(I_{\mathrm{2c}}\) are the critical current of the SIs and sFS side, respectively. The fact that the transport properties are dominated by the SIs side indicates that \(I_{\mathrm{1c}}\)  is much smaller than  \(I_{\mathrm{2c}}\) and thus the phase drop $\psi$ is negligible \begin{math} \left ( \sin \psi \approx  I_{1c}/ I_{\mathrm{2c}} \sin \varphi \ll 1  \right ) \end{math}.  
As shown in the Supplementary Notes 4, the following equation for \(\varphi\) is derived: 
\begin{equation}\label{Relazione_phi'_Meff}
\partial_{\mathrm{x}} \varphi  = \frac{2\pi }{\Phi_{0}}  \left ( 2\mu_{\mathrm{0}} H\lambda_{\mathrm{L}}+ \mathcal{M}_{\mathrm{eff}}  \right ),
\end{equation}
where \(\mathcal{M}_{\mathrm{eff}}\) is the effective magnetic moment considering the geometry of our device:
\begin{equation}\label{M_eff}
 \mathcal{M}_{\mathrm{eff}}  = \mu_{\mathrm{0}} M_{\mathrm{F}} d_{\mathrm{F}} \left (  1 - \gamma  \right ) e ^{ -\theta_{\mathrm{S}}},
\end{equation}
where \(\theta_{\mathrm{S}} = d_{\mathrm{s}}/ \lambda_{\mathrm{L}} \sim 0.3\), in our junction. Hence, in analogy with Eq. (\ref{flux_dahir}):
\begin{equation}\label{Flux_spin_polarization}
\Phi =  2\mu_{\mathrm{0}}H2R\lambda_{\mathrm{L}} + \mu_{\mathrm{0}}M_{\mathrm{F}}2Rd_{\mathrm{F}}  \left [ 1 - \gamma \right ] e ^{ -\theta_{\mathrm{S}}}.
\end{equation}
Eq. (\ref{Flux_spin_polarization}) implies that for a series connection of a tunnel SIs and a ferromagnetic sFS junction with  \(d_{\mathrm{s}} < \lambda_{\mathrm{L}}\) the shifts of the magnetic field patterns are related to $\gamma$  and we can thus derive its temperature dependence, reported in Fig. \ref{temperature_dependence_induced_magnetization}c. It turns out that the lack of magnetic hysteresis is related, within the experimental errors, to an almost full spin screening regime, while by increasing the temperature $\gamma$ is reduced and the magnetic patterns show a progressive hysteretic behavior.
More importantly, this junction layout allows to disentangle spin polarization phenomena from uncontrolled anomalies in magnetic field patterns of MJJs, such as those due to domain structure of the F-layer, thus providing an unambiguous evidence of the spin polarization of the Cooper pairs at the S/F interface.

For Nb/Py proximity-coupled system, for which \(J /  \Delta \sim 10\) (Ref. [\citenum{Robinson2006}]), the full spin screening is expected to occur at low temperatures if \(\varepsilon_{\mathrm{b,F}}/ \Delta\) is larger than 10 (Fig. \ref{temperature_dependence_induced_magnetization}b), which is consistent with our experiment and with the measured values of  \(R_{\mathrm{b}}\), entering the estimation of  \(\varepsilon_{\mathrm{b,F}} = \hslash D_{\mathrm{F}}/({R_{\mathrm{b}}\sigma_{\mathrm{F}}d_{\mathrm{F}}})\). For our F-films, \(R_{\mathrm{b}}\) is of  the order of magnitude of f\(\Omega\)m$^{2}$ as for MJJs with Nb/Py interface \cite{Bell2005}.  In the case of SIsFS JJs,   \(\mathcal{M}_{\mathrm{eff}}\)  is not directly coupled to the phase difference $\varphi$ across the SIs. Nevertheless, the broadening of the central peak, with a factor of about 2.5 at low temperatures, is evident and represents the second hallmark of the IPE. Indeed, the width of central peak decreases by increasing the temperature and becomes consistent  with the geometric expectations and the magnetization reversal of our Py dots at \textit{T }= 6 K (see Supplementary Figure 3b), when the IPE is negligible. 

Further consistency is given by SIsFS JJs with a thickness of the s-interlayer of 10 nm. In this case, since \(d_{\mathrm{s}} \sim \xi_{\mathrm{s}}\) the inner IsF trilayer acts as a single Josephson barrier \cite{BAKURSKIY2013,BAKURSKIY2013_PRB}, thus resulting in \textit{I}\(_{\mathrm{c}}\)\textit{R}\(_{\mathrm{N}}\) reduced by about two orders of magnitude, of the order of tens of \(\mu\)V at \textit{T} = 10 mK. In this regime,  as shown in Supplementary Figure 6a, we have again observed zero-centered  \(I_{\mathrm{c}}\) curves and a broadening of the central peak even larger as expected for a smaller value of \(\theta_\mathrm{s}\). In contrast, in samples with a 14 nm-thick PdFe, an ordinary hysteresis of the \(I_{\mathrm{c}}\) curves is restored, as expected by the theory for weak and thick F-interlayer\cite{CARUSO2018}. Finally, as shown in the Supplementary Notes 5, the contribution due to the spontaneous Meissner supercurrent in response to the vector potential in F at the S/F interface can be neglected \cite{Mironov2018}. Therefore, we can conclude that the origin of the temperature dependence of the \(I_{\mathrm{c}}\) curves has to be ascribed to the spin polarization of the Cooper pairs at the S/F interface.  

\begin{figure} [h]
   \centering
    \includegraphics[width = 9 cm]{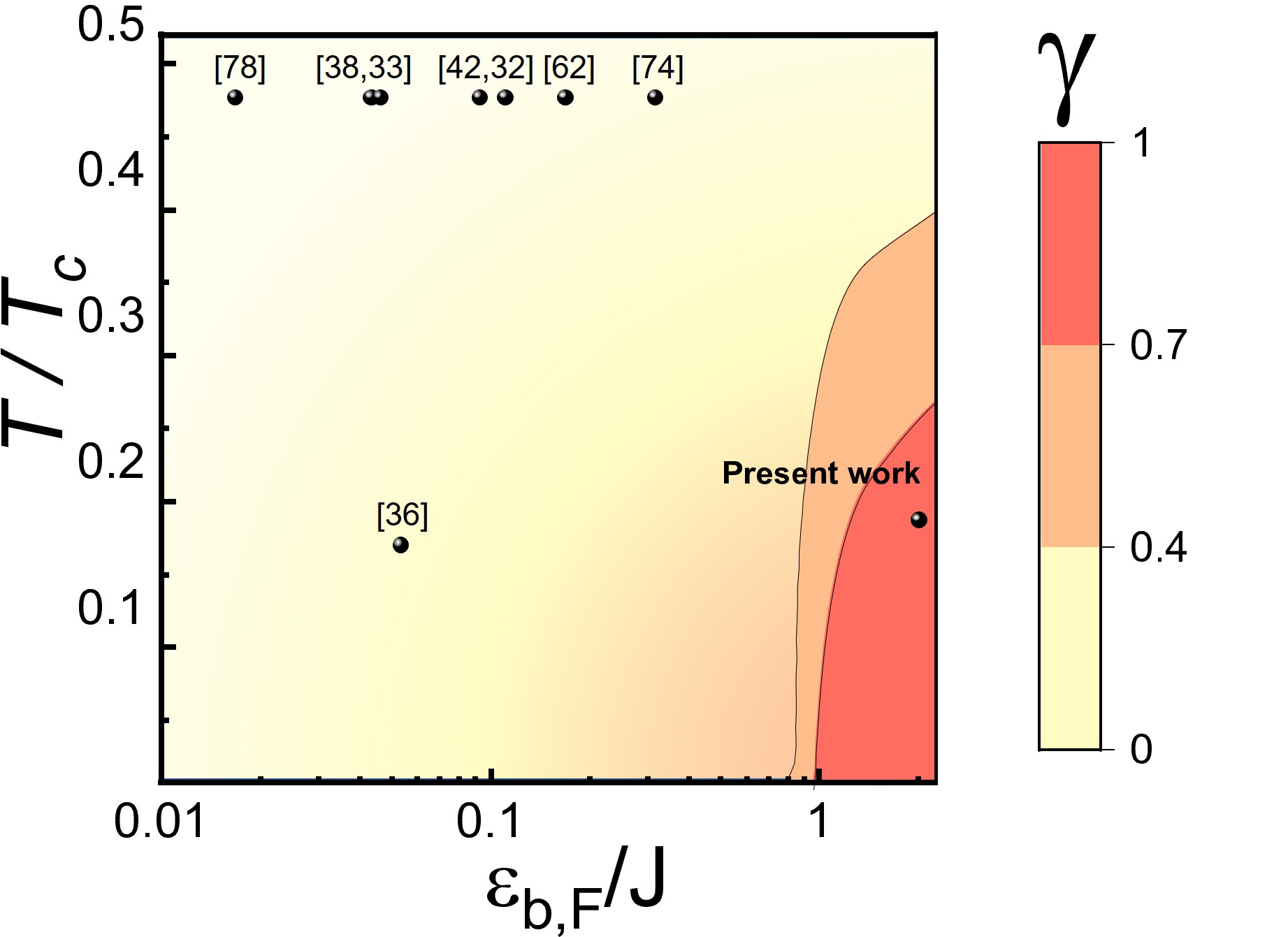}   
 \caption{\textbf{Phase diagram for the dimensionless magnetization $\gamma$.}  Phase diagram for the dimensionless magnetization $\gamma$, which depends on the ratio \(\varepsilon_{\mathrm{b,F}} /J\) and \(T/T_{c}\).  At low temperatures, a strong polarization of the S/F interface is expected for \(\varepsilon_{\mathrm{b,F}} /J \geq 1\) (pink background). In contrast, in  the typical experimental conditions reported in literature (black dots labeled with the corresponding reference), the magnetization is weakly induced (light-yellow background). For more information on the parameters extracted from literature, we refer to Supplementary Table 2.} \label{comparison_with_the_literature}
\end{figure}

The phase diagram reported in Fig. \ref{comparison_with_the_literature} condenses the various spin screening regimes according to the main physical parameters, i.e., temperature \textit{T} and characteristic energy scales of the ferromagnet \textit{J}  and \(\varepsilon_{\mathrm{b,F}} \). The following general conclusions can be inferred:  i) Low temperatures are required to observe spin screening effects. To date, the \(I_{\mathrm{c}}\) measurements have been mostly performed at helium-liquid temperature to demonstrate the functionality of MJJs as switchable elements for digital electronics \cite{SOLOVIEV2017} and for spintronic devices \cite{ECHRING2015}. At that temperature, the induced magnetization is significantly reduced and thus the effects of the spin polarization become hard to be isolated. ii) The strong polarization limit is characterized by large values of  \(\varepsilon_{\mathrm{b,F}} \), which can be achieved by employing strong and thin ferromagnet directly coupled to the S layer.  In contrast, the use of buffer layer prevents  the polarization of the S/F interface \cite{Bell2005,Bell2004_APL, GLICK2017,BAEK2014, QADER2014,Martinez2016, Niedzielski2018}, while weak and thick ferromagnets suppress the value of $\varepsilon_{\mathrm{b,F}}$ and thus $\gamma$ even at low temperature \cite{sellier2003,Oboznov2006,GLICK2017,kontos2002,BOL'GINOV2012}. 

Finally, these findings are not only important steps forward in improving the description and understanding of proximity-coupled systems, but also in implementing these MJJs for quantum devices. At the operating temperature of quantum circuits, the IPE can emerge (Fig.  \ref{comparison_with_the_literature}) and lead to a significant modification of the functioning of the overall device. For the ferro-trasmon, the screening of the F magnetic moment and the resulting lack of the hysteresis represent a drawback since the latter prevents the tuning of the qubit frequency \cite{Ahmad_22PRB,ferro-trasmon_path}. We have faced this issue by realizing SIsFS JJs based on Al technology and with Py as F-layer \cite{Vettoliere22}. In this case, a thin natural AlO$_{\mathrm{x}}$ barrier forms at the S/F interface and decouples the s and F layers: as a result, the transport properties of SIsFS junctions are not affected by the presence of the ferromagnet, while the spin polarization of the S/F interface is weakly induced, resulting in an ordinary hysteresis of the \(I_{\mathrm{c}}(H)\) curves even at \textit{T} = 10 mK (as shown in Supplementary Figure 6b).  This experimental observation is in agreement with the theoretical prediction that a highly transparent S/F interface is a key factor to observe full screening at low temperatures  (Fig.  \ref{comparison_with_the_literature}). Moreover, this is also a proof that, when the conditions of the IPE are not realized, standard behavior of the magnetic field patterns are fully recovered. Finally, as addressed in the Supplementary Note 6 and show in Supplementary Figure 7, the  measurements of the \(I_{\mathrm{c}}(H)\) curves by varying the temperature allow to identify the presence of the spin polarization of the Cooper pairs even when different magnetic interactions coexist at the S/F interface. 
\section*{Conclusions}
In conclusion, by exploring a new region of the \(\gamma(T, \varepsilon_{\mathrm{b,F}} /J)\) phase diagram of MJJs, we have demonstrated the full screening of the F-magnetic moment in SIsFS tunnel JJs. The Josephson effect, being sensitive to phase space variation even on the scale of nanometers, gives -because of intrinsic nature- macroscopic information mediated at the nanoscale. Our experiment establishes another milestone in the study of the rich physics of the S/F interface and can inspire the search for new hybrid orders in non-conventional systems. A deep understanding and control of proximity junctions is also fundamental for the design of the S/F interfaces and further developments  for digital and quantum superconducting electronics.
\section*{Methods} \zlabel{methods}
\subsubsection*{Sample fabrication}
A Nb-Al/AlO$_{\mathrm{x}}$-Nb trilayer has been deposited onto oxidized 3-inches Si wafer by using d.c. magnetron sputtering in ultra-high vacuum system. The base and the top electrodes consist of Nb films having a thickness of 200 nm and 40 nm respectively, deposited at rate of 1.2 nm s$^{-1}$. The intermediate Al layer has been deposited at a small rate of 0.7 nm s$^{-1}$ to obtain a film thickness of 7 nm, which afterwards is exposed to dry oxygen for 1 h to form the AlO$_{\mathrm{x}}$ tunnel barrier. The trilayer has been patterned using optical lithography and lift-off procedure, while the junction areas have been obtained by a selective anodization process together with a further insulation by SiO$_{2}$ deposition.  Then the wafer has been diced into 10 x 10 mm$^{2}$ chips and a soft Ar ion etching has been used to remove about 10 nm of Nb oxide layer before depositing the ferromagnetic layer by lift-off technique.  The 3 nm-thick Ni$_{80}$Fe$_{20}$ layer layer has been sputtered by a magnetron source at a rate of 0.7 nm  s$^{-1}$. Finally, a 400 nm top Nb counter electrode has been deposited by a further d.c. sputtering and lift-off processes obtaining the overall SIsFS structure, i.e, a Superconductor/ Insulator/ thin superconductor/ Ferromagnet/ Superconductor stacked multilayer \cite{JAP}.
\subsubsection*{Measurements set-up}
 
The SIsFS Josephson junctions (JJs) have been measured by thermally anchoring the samples to the mixing chamber of a Triton dry dilution refrigerator provided by Oxford instruments, with customized low noise filters anchored at different temperature stages \cite{Caruso_PRL, Ahmad_22}. The junction is current-biased with a low frequency current ramp (approximately 11 Hz) using a waveform generator in series with a shunt resistance, while the voltage across the junction is measured using a battery-powered differential amplifier.  Magnetic field in the plane of the junction can be applied using a NbTi coil \cite{Caruso_PRL, Ahmad_22}.  
Concerning the measurements of magnetic field pattern, the first measurements have been performed at temperature \textit{T} = 10 mK and then the curves have been acquired by increasing the temperature. In order to avoid trapping flux in the superconducting Nb layers, we have always warmed the sample to the next temperature in zero field.

\section*{Data availability}
The data that support the findings of this study are available from the corresponding
authors upon reasonable request.

\section*{Code availability}
Codes written for and used in this study are available from the corresponding authors upon reasonable request. 

\typeout{} 
\bibliography{sample} 

\section*{Acknowledgements}
The authors are grateful to Carla Cirillo and Antonio Leo for their support in the evaluation of the coherence lengths. This work has been supported by the Pathfinder EIC 2023 project "FERROMON-Ferrotransmons and Ferrogatemons for Scalable Superconducting Quantum Computers", the PNRR MUR project PE0000023-NQSTI, the PNRR MUR project CN-00000013-ICSC, the project SuperLink—Superconducting quantum-classical linked computing systems, call QuantERA2 ERANET COFUND, CUP B53C22003320005, the project “SQUAD-On-chip control and advanced read-out for superconducting qubit arrays,” Programma STAR PLUS 2020, Finanziamento della Ricerca di Ateneo, University of Napoli Federico II, and the project “EffQul- Efficient integration of hybrid quantum devices”-Ricerca di Ateneo Linea A, CUP: E59C20001010005. H.G.A., D.Ma. (Davide Massarotti), D.Mo. (Domenico Montemurro) and F.T. thank SUPERQUMAP project (COST Action CA21144). 

\section*{Author contributions}
R.S., G.A. and Da.M. conceived the experiments; A.F.V. worked on the theoretical model; R.S., H.G.A, L.D.P. and R.F. carried out the measurements; R.S., A.F.V., H.G.A, L.D.P. and R.F. worked on the data analysis; R.S., A.V., C.G. and L.P. designed and realized the junctions; G.P. P. and  F.T.  contributed to the fundings; R.S. and Da.M. co-wrote the original draft; R.S., A.F.V., H.G.A, Do.M., L.P., G.P. P.,  F.T., G.A. and Da.M. review and edit the paper. All authors discussed the results and commented on the manuscript.
\section*{Competing interests}
The authors declare no competing interests.


\end{document}